\documentclass[aps,showpacs,twocolumn]{revtex4}
\usepackage{graphicx}
\usepackage{amsmath}
\usepackage{amsfonts}
\usepackage{pifont}
\usepackage{mathrsfs}

\begin{document}
\newcommand{\bstfile}{aps} 
\newcommand{\bibs}{IntTheoryRefs,BibFile}
\title{Experimental tests of the correlated chromophore domain model of self-healing in a dye-doped polymer}
\author{Shiva K. Ramini\footnote{Current address: Wyatt Technology Corporation, 6300 Hollister Ave,
Santa Barbara, CA 93117-3253}} 
\address{Department of Physics and Astronomy, Washington State University \\ Pullman,
Washington  99164-2814}
\author{Sheng-Ting Hung}
\address{Department of Physics and Astronomy, Washington State University \\ Pullman,
Washington  99164-2814 \\}
\address{Department of Chemistry \\ Katholieke Universiteit Leuven \\ Celestijnenlaan 200D \\ B-3001 Leuven, Belgium}
\author{Mark G. Kuzyk}
\address{Department of Physics and Astronomy, Washington State University \\ Pullman,
Washington  99164-2814}
\date{\today}

\begin{abstract}
Temperature dependent photodegradation and recovery studies of Dipserse Orange 11 (DO11) dye dissolved in poly(methyl methacrylate) and polystyrene polymer hosts are used as a test of the recently proposed correlated chromophore domain model.\cite{ramin12.01} This model posits that dye molecules form domains or aggregates. The nature of aggregation or how it mediates self healing is not yet well understood. In this paper we present qualitative evidence that supports the hypothesis that the dye molecules undergo a change to a tautomer state with higher dipole moment and hydrogen bond with the amines and keto oxygens of the polymer.  Groupings of such molecules in a polymer chain form what we call a domain, and interactions between molecules in a domain make them more robust to photodegradation and mediate self healing.
\end{abstract}


\maketitle

\vspace{1em}

\section{Introduction}

Structural damage and degradation of a polymer is usually associated with cracking.  Mitigating damage or developing methods to promote healing in polymeric materials after cracking is an active area of research motivated by its practical utility.  White and coworkers reported on a structural polymeric material with the ability to autonomically self-repair cracks.\cite{white01.01} Such polymers incorporate a microencapsulated healing agent that is released in the cracking process with polymerization being triggered by contact of a catalyst with the healing agent, thus bonding the crack faces. White observed as much as
75\% recovery in toughness.

Our work presented here is different in two regards.  First, the sample is a dye-doped polymer rather than a neat polymer and the degradation process is through optically-induced burning, so chemical changes are induced rather than solely mechanical/structural damage - though cracking can accompany burning.  The dopant molecules thus mediate the phenomena.  The degree of damage is observed using optical techniques, the simplest of which is the detection of a color change.  Secondly, the healing process is a microscopic one, originating at a molecular level that we believe involves a cooperative process of aggregates of molecules.  The materials of interest to our work have applications as optical materials where photodegradation is a common cause of optical and optoelectronic device failures, either as catastrophic failure or a slow deterioration of the performance.\cite{dyuma92.01,popov.98.01,zhang98.01,galvan20.01}

Though originating at the microscopic level, self healing in dye-doped polymers also leads to macroscopic healing.  For example, DesAutels et al have shown that holes drilled into a dye-doped polymer with high-fluence laser pulses heal, i.e. contract over time, in the presence of the dye.\cite{desau09.01}  Without the dopant molecules, the holes enlarge and develop ragged edges.  Analogously, self healing is not observed when molecules are in solution in a monomer solvent,\cite{howel04.01} but only in solid solution with the polymer as the host.\cite{howel02.01} Thus, the healing process appears to be a cooperative one between the solute dyes and the polymer.

The discovery of reversible photodegradation was a serendipitous one. Peng first observed recovery of fluorescence after photodegradation of a dye-doped polymer optical fiber,\cite{Peng98.01} and, Howell and Kuzyk discovered that Disperse Orange 11 (DO11) dye dissolved in poly (lethyl methacrylate) (PMMA) polymer undergoes full recovery after photodegradation using Amplified Spontaneous Emission (ASE) to monitor the process.\cite{howel01.01,howel02.01} Embaye et al showed that molecules photo degrade to a long-lived damaged state, which requires the polymer host.\cite{embay08.01} The process was modeled using two higher-energy states, one the damaged state, that both decay back to the undamaged state.  These states were probed with linear absorption spectroscopy. Ramini et al further investigated the underlying mechanisms and showed that dye molecules when embedded in polymer matrix appear to form domains, and cooperative effects within a domain lead to both protection from photo-damage and recovery.\cite{ramin12.01}

There have been many proposed mechanisms for recovery, including photo reorientation (orientational hole burning), photo-induced isomerization, and diffusion.  All of these have been shown not to be responsible.\cite{embay08.01,ramin11.01} Our present working hypothesis is that domains are involved and that the distribution of domains are governed by a condensation process.\cite{ramin12.01} Whilst the energetics of the process are consistent with a tautomer of the DO11 molecule hydrogen bonding to the polymer backbone,\cite{kuzyk12.01} the same aggregation model may be consistent with the formation of a twisted charge transfer structure.\cite{dirk12.01}  For the purposes of this paper, the distinction between the two are unimportant.

In the present work, we provide additional tests of the correlated chromophore domain model using temperature-dependent measurements. Past studies focused on poly (methyl methacrylate) (PMMA) polymer as the host.  Here, we also test a random copolymer of PMMA and polystyrene (PS) to investigate the effect of changing the polymer host. The new measurements provide further support for our model.

\section{Chromophore Correlation Model-Brief overview}

In our earlier paper\cite{ramin12.01} we proposed a model of reversible photodegradation of Disperse Orange 11 (DO11) dye doped in Poly(methyl methacrylate) (PMMA) that has as its basis localized correlated regions of molecules, which we call domains.  These domains have not been physically characterized; rather, indirect evidence supports the hypothesis.  Here we provide additional evidence.

To summarize the model, for a domain of size N, the undamaged population decays in proportion to the intensity and the population of available undamaged molecules in the domain, $n$, while recovery is in proportion to the number of damaged molecules, $N-n$.  Empirically, we have determined that the bulk material behaves in a way such that the decay rate is slower for larger domains,  and the recovery rate is accelerated in proportion to the number of undamaged molecules in a domain. This behavior is quantified by the phenomenological model of a single domain given by
\begin{equation}\label{InteractingRates}
\frac {d n } {dt} = \beta n (N - n) - \frac{\alpha I} N n
\end{equation}
where $\alpha$ is the rate per unit intensity at which molecules decay, $\beta$ is the recovery rate, and $I$ is the intensity of the light that damages the sample.  This model was shown to fit the data well for DO11 dye doped in PMMA polymer as a function of concentration, time, and intensity.\cite{ramin12.01}

Integrating Equation \ref{InteractingRates} yields
\begin{equation}\label{RecoverN}
n(t) = \frac { \left( N - \alpha I / \beta N \right) n_0} {n_0 + \left( N - n_0 - \alpha I / \beta N  \right) \exp \left [ - \left(N \beta - \alpha I/N \right) t\right] } ,
\end{equation}
where $n_0$ is the initial undamaged population at $t=0$.

When the pump intensity is turned off ($I=0$), the population recovers according to
\begin{equation}\label{InteractingRates2}
\frac {d n } {dt} = \beta n (N - n).
\end{equation}
With $n(t_0)$ as the undamaged population right at the time the pump laser is turned off, the solution to Equation \ref{InteractingRates2} can be written as,
\begin{equation}\label{RecoverN2}
n = \frac { N }{1 + \left[ \frac N {n(t_0)}- 1\right] e^{\left( -N \beta t\right)}} .
\end{equation}
\begin{center}
\begin{table}
\caption{Parameters determined for DO11 in PMMA using an average pump intensity of $I_p = 0.202 W/cm^2$.\label{tab:results}}
  \begin{center}
  \begin{tabular}{c  c  c }
  \hline
    $\alpha (min^{-1}W^{-1}cm^2)$& $\beta (10^{-4}min^{-1}) $ & $\delta \mu (eV)$  \\
    Decay Rate & Recovery Rate & Free energy/molecule \\ \hline\hline
    $7.09 (\pm 0.13)$ & $3.22 (\pm 0.26)$ & $0.29 (\pm 0.01)$ \\ \hline
  \end{tabular}
  \end{center}
\end{table}
\end{center}

The above calculation is for a single domain.  The observed behavior of a sample originates in the average behavior over the domains.  We assume that the domains are distributed according to a condensation model.  It is energetically favorable for a domain to form -- the more molecules in a domain, the lower the energy.  However, the temperature, $kT$ breaks up the domains.  At zero temperature, one would expect a single large domain, and at infinite temperature, an equal number of domains of each size.  Thus, since the average domain size is larger at low temperature, the recovery rate in this model would increase as the temperature is lowered.  This behavior is opposite to a barrier model, in which elevated temperature would accelerate healing.  Thus, the temperature dependence of photodegradation and recovery is a critical test of the theory.

Formally, the domain distribution is calculated by minimizing the free energy of the system. After considerable calculation, the distribution is found to be given by\cite{ramin12.01}
\begin{align}\label{domainno2}
\Omega(N)&= \frac 1 z \left[\frac{(1+2\rho z)-\sqrt{1+4\rho z}}{2\rho z}\right]^N,
\end{align}
where $z=\exp\left(\frac{\delta \mu}{kT}\right)$. $\Omega(N)$ is the number density of domains of size $N$, $\rho$ is the average number density of molecules in the sample, $\delta \mu$ is the free energy advantage per molecule -- which is the energy released when a molecule is added to a domain, $T$ is the temperature, and $k$ is the boltzmann constant.

Using Equations \ref{RecoverN} and \ref{domainno2}, the mean number of undamaged molecules in a domain, $\overline{n}$, is thus given by the ensemble average,
\begin{align}
\nonumber\overline{n}(t;\rho ,T,I,n_0)&= \sum_{N=1}^{\infty}n(t;N,I)\Omega(N;\rho, T)\\
 &\approx\int_1^{\infty} n(t)\Omega(N)dN \label{IntegrateDistribution}
\end{align}
The signal observed from optical measurements is related to this mean population.

In our earlier paper, we used the concentration/intenisity-dependent data to determine the material parameters. As a test of the model, we used one set of parameters determined from population decay data to predict the recovery dynamics at several dye concentration and showed that experimental observations agree with the predictions.  Table \ref{tab:results} summarizes the values of the three parameters in the model which characterize a particular composite material, in our case, DO11 dye in PMMA polymer.

\section{Temperature Dependent Measurements}

\subsection{Experimental Setup}

We use amplified spontaneous emission (ASE) as a sensitive probe of the population of undamaged molecules by taking advantage of the fact that undamaged molecules lase strongly while damaged molecules do not.  The sensitivity of this measurement originates in the nonlinearity of the process.  Small changes in the population due to damage yield large changes in the ASE intensity.  The experimental configuration is essentially a strong laser beam that both damages the sample and excites the undamaged molecules, which then emit coherent light.  The ASE light, which is red, is emitted perpendicular to the direction of the green pump beam.  Thus, spectral and spatial filtering can be used to sperate the red signal from the green pump beam.

\begin{figure}
\includegraphics
{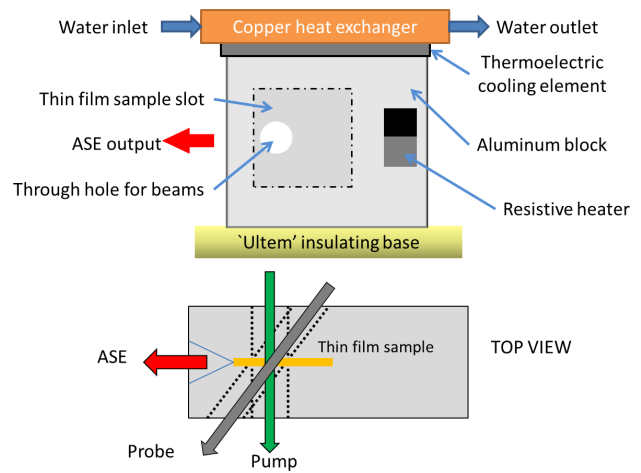}
\caption{Custom sample holder and heating/cooling system used for temperature dependent studies.}
\label{fig:chamber}
\end{figure}

The sample chamber must provide a means for heating and cooling the sample while passing the pump light and collecting the ASE light. There is also the option of simultaneously measuring changes in absorbance by allowing a probe beam to overlap the burned area generated by the pump. A schematic cross sectional view of the chamber is shown in the Figure \ref{fig:chamber}.

The sample holder is made of aluminum due to its good thermal conductivity and is cooled using a thermoelectric(TE) element. The TE element works on the principle of the Peltier effect. When current is passed through the terminals, it extracts heat from one side of the element and deposits it out the other side. The TE element is placed on top of the sample holder so that the cooling side faces down and in contact with the sample holder using thermal conductive paste. A water-cooled copper heat exchanger is placed on top (hot side) of the TE element to extract heat out of the system. A thick film resistor is attached on the side to heat the sample holder.

Two through-holes are drilled into the aluminum block, one normal to the sample to pass the pump laser and the other one at a $30^0$ angle to the normal for probing with white light. The center axis of the two holes overlap at the sample allowing the white light and pump laser to cross each other. An conical opening is made in the side of the holder for emitted ASE to be detected. The sample holder is placed on an insulating base made of `Ultem' plastic to avoid heat exchange with the environment. This plate is attached to an xyz-translation stage (not shown). The whole setup is covered with a plexi-glass enclosure to reduce heat exchange with the environment due to convection. Sapphire windows are installed in the enclosure walls where the beams enter and exit the enclosure. Desiccant is placed inside the enclosure to avoid water condensation at lower temperatures. A K-type thermocouple is inserted in the holder through a drilled hole to measure the instantaneous temperature near the sample. Thermoelectric element, heating resistor, and thermocouple are connected to an Omega CN7833 temperature controller which is interfaced with a computer.

\subsection{Results and Discussion}

\subsubsection{Temperature Dependence}

\begin{figure}
\includegraphics{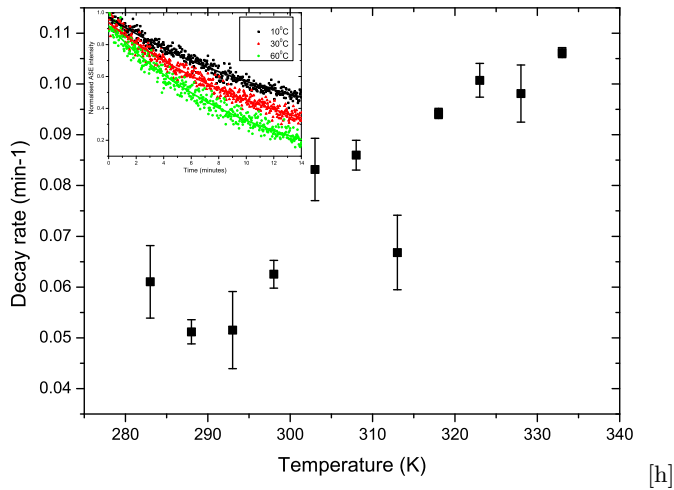}[h]
\caption{Decay rates estimated from single exponential fits to the ASE decay data(inset) as a function of sample temperature.}
\label{fig:tempexpdecay}
\end{figure}

\begin{figure}
\includegraphics{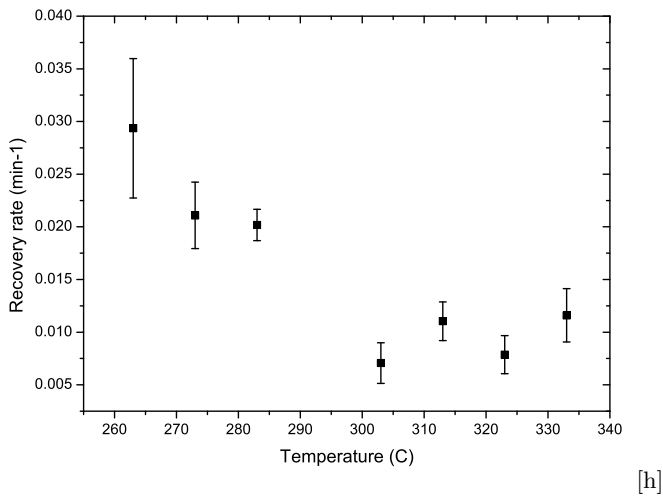}[h]
\caption{Estimated recovery rates from single exponential fits to the recovery data as a function of sample temperatures.}
\label{fig:tempexprec}
\end{figure}
A 10g/l DO11 dye mass to PMMA volume sample is used for the temperature-dependent experiments.  This concentration is chosen because it produces good ASE signal and has a domain distribution of large enough average size to exhibit good self healing.  The decay and recovery data is obtained at several temperatures.

Rather than using the full model, we initially estimate the decay and recovery rates using a single exponential fit.  Exponential functions yield a good fit to the data, and provide a good estimate of the decay and recovery rates.  Figures \ref{fig:tempexpdecay} and \ref{fig:tempexprec} show the decay and recovery rates so determined. Each data point at each temperature is the weighted average of several runs. The data clearly shows that recovery is slower at high temperature and is consistent with the domain model and eliminates the barrier hypothesis.

\begin{figure}
\includegraphics{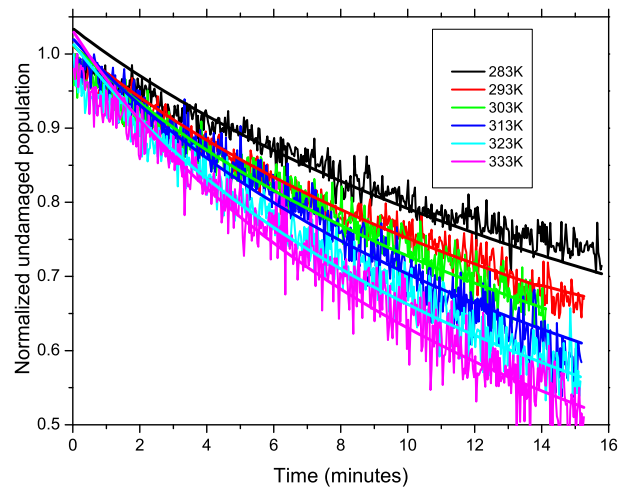}
\caption{Population decay upon irradiation at several temperatures along with the predictions from the Chromophore Correlation Model.}
\label{fig:tempdecay}
\end{figure}

Figure \ref{fig:tempdecay} shows population decay curves at six different temperatures. The data is plotted along with the chromophore correlation model predictions (lines). The predictions from the model are calculated using the parameters in Table \ref{tab:results}. No adjustable parameters are used except a global normalizing constant at the beginning of Equation \ref{IntegrateDistribution} which takes into account the detector efficiency and alignment changes from run to run.

The trend in Figures \ref{fig:tempexpdecay} and \ref{fig:tempexprec} can be explained using the chromophore correlation model as follows. As the temperature increases, thermal motion, or entropy, competes with the tendency for molecules to form domains leading to lowering of average domain size as explained in the literature\cite{ramin12.01}. If domains are caused by a condensation-like process, the domain size grows as the temperature is decreased and dissociate when the the temperature in increased. In this picture, an increase in the temperature leads to a decrease in the average domain size. With smaller domains, there are fewer healthy molecules to coax the damaged ones to heal.  Since the decay rate is inversely proportional to the domain size, the net effect is that as the temperature is increased, the decay rate of the undamaged population increases, consistent with our observation in Figure \ref{fig:tempexpdecay}. The recovery rate also decreases as the domain size decreases. This leads to a slower recovery rate also consistent with our observations. Hence, the model correctly predicts the temperature dependence of self healing.

\subsubsection{Nature of the Domains}

\begin{figure}
\includegraphics{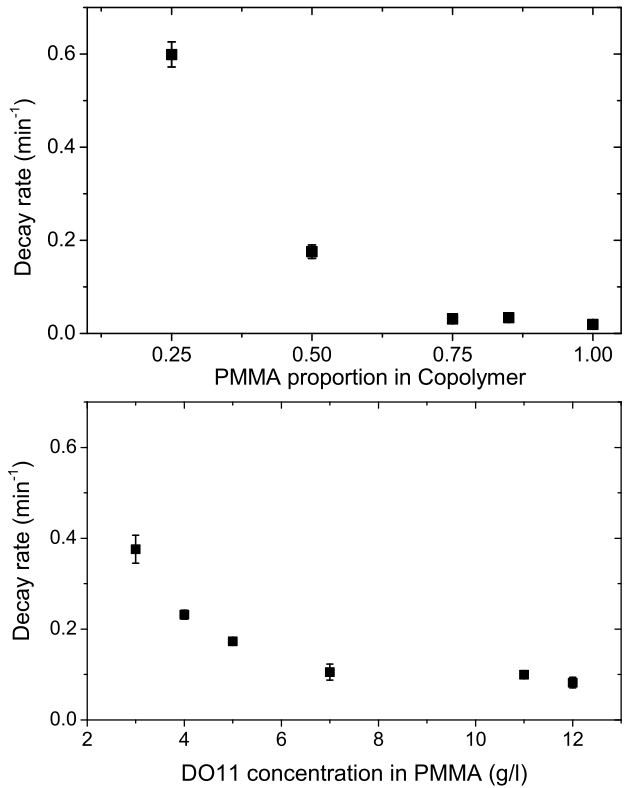}
\caption{(top) DO11 population decay rate as a function of PMMA content in PMMA/PS copolymer while dye concentration is kept constant. (bottom) DO11 population decay rate as a function of dye concentration in pure PMMA polymer.}
\label{fig:PSvsConc}
\end{figure}

While the chromophore correlation model correctly predicts the behavior of the dye-polymer system, the nature of aggregation and the role of the polymer is yet to be understood. Based on the fact that recovery of ASE in DO11 is not observed in solution in liquid MMA monomer, the solid polymer clearly plays an important role. Kuzyk et al discussed several possible sources of aggregation and the role of the polymer.\cite{kuzyk12.01} The hypothesis that interactions between a tautomer of the DO11 molecule with the PMMA polymer chain is the origin of the interaction that forms a domain is based on the observation that the parameter $\delta \mu$ is about the same as the hydrogen binding energy calculated for the tautomer interacting with PMMA.\cite{kuzyk12.01} It is also possible that it is a TICT state, but its energetics are less well known or inaccurate to test this hypothesis.\cite{dirk12.01}

The model predicts that the average domain size naturally decreases as the concentration is decreased.  Another method for decreasing the domain size is to form a copolymer in which the second component does not hydrogen-bond with the DO11 tautomer.  Polystyrene is such a polymer. To test this hypothesis, we measured the population decay rates in samples of dyes of fixed concentration ($9 g/L$) in PMMA-PS copolymer.  By varying the weight ratio of styrene to MMA, the domain size can be controlled.  As a point of reference, we can compare this data with dye concentration-dependent data in PMMA.

Figure \ref{fig:PSvsConc} shows the data. Reducing the PMMA content in the copolymer behaves in a similar way to reducing the dye concentration in a pure PMMA sample. This behavior is consistent with our model and the domain picture.  The model of domain formation in a copolymer will need to be refined in order to better test the hypothesis quantitatively. Also, additional experiments that measure both decay and recovery; and, better estimate the various parameters are needed in order to strengthen our case.

\section{Conclusions}
We have shown that the correlated chromophore domain model predicts all the observations of decay/recovery of a dye-doped polymer as a function of time, intensity, concentration, and temperature using only three {\em fixed} parameters. Most importantly, the temperature-dependent data -- which shows accelerated healing at decreased temperatures, eliminates the barrier hypothesis and is consistent with our domain picture of the degradation and healing process.

Also tested is the hypothesis that domain formation originates in hydrogen-bonding of the DO11 molecules with the PMMA polymer chain by evaluating copolymers in which one of the components does not provide bonding sites.  Complementary evidence that supports the domain picture is the observation that the decay dynamics as a function of PMMA fraction is similar to the decay dynamics as a function of dye concentration.  In both cases, the number of dye molecules in proximity to PMMA polymer chains is the same, so interactions between the two are responsible.  Since hydrogen bonding of the DO11 molecule with the polymer is the most pronounced difference between PMMA and PS, and the energy parameter $\delta \mu$ is consistent with the hydrogen bonding energy between a DO11 molecule and PMMA polymer,  we propose that hydrogen bonding is the most obvious mechanism.

Hydrogen bonding as the mechanism of domain formation does not explain the healing mechanism.  It is likely, however, that healing originates in and is accelerated by interactions that make healing energetically more favorable.  Clearly, a more quantitative analysis is required with supplemental data to make possible a more definitive conclusion.

The concept that a material would exhibit such complex behavior without intentional design by the experimenter is an interesting one.  Though self healing is a process with great practical utility, it is intriguing that nature has been kind enough to provide an inherently smart material system that appears to behave in a way contrary to most others; it mediates recovery in a world in which irreversible damage is the norm.  Further advances in understanding the physics underlying this phenomena will surely enable new applications that require materials to withstand high light intensities; and, may lead to new physics.

\section{Acknowledgements}
We thank the Air Force Office of Scientific Research (FA9550-10-1-0286) for their generous support.
\bibliography{\bibs}

\end{document}